\documentstyle[preprint,aps,prb]{revtex}

\begin{document}
 
\newcommand{\beq}{\begin{equation}}
\newcommand{\eeq}{\end{equation}}
\draft

\newcommand{\U}{{\cal U}} 
\def\G#1#2{{\cal G}^0_\Delta(#1,#2;\beta)} 
\def\GE#1#2{{{\cal G}^0_\Delta}_E(#1,#2;\beta)} 
\newcommand{\K}{{\cal K}^0_\Delta}
\newcommand{\g}{g^0_\Delta} 
\newcommand{\Ib}{\int_{-\beta/2}^{\beta/2}}
\def\intq#1{\int \frac{d^4#1}{(2\pi)^4}}
\def\intt#1{\int \frac{d^3#1}{(2\pi)^3}}
\def\intf#1{\int \frac{d#1_0}{2\pi}}
\newcommand{\bsig}{\mbox{\boldmath $\sigma$}}
\newcommand{\btau}{\mbox{\boldmath $\tau$}}
\def\bfm#1{{\mbox{\boldmath $#1$}}}
\newcommand{\Tr}{{\rm Tr}}
\newcommand{\llan}{\left<}
\newcommand{\rran}{\right>}

\bibliographystyle{unsrt} 

\title
{\bf Path Integral Variational Methods \\
for Strongly Correlated Systems}
\author{T.S. Walhout$^{ab}$, R. Cenni$^c$,
A. Fabrocini$^d$ and S. Fantoni$^{ab}$}
\address{ $^a$ Interdisciplinary Laboratory,
 International School for Advanced Studies \\
 and Istituto Nazionale di Fisica Nucleare - sez. di Trieste\\
 Via Beirut 2 - 34014 Trieste (Italy)}
\address{$^b$ International Center for Theoretical Physics\\
 Strada Costiera 11 - 34014 Trieste (Italy)}
\address{$^c$ Istituto Nazionale di Fisica Nucleare - sez. di Genova\\
 and Dipartimento di Fisica - Universit\`a di Genova\\
 Via Dodecaneso 33 - 16146 Genova (Italy)}
\address{$^d$ Dipartimento di Fisica - Universit\`a di Pisa\\
 and Istituto Nazionale di Fisica Nucleare - sez. di Pisa\\
 Piazza Torricelli 2 - 56100 Pisa (Italy)}

\preprint{IFUP-TH 24/96}
\maketitle

\begin{abstract}  
We introduce a new approach to highly correlated systems
which generalizes the Fermi Hypernetted Chain and
Correlated Basis Function techniques. While the latter
approaches can only be applied to systems for which a
nonrelativistic wave function can be defined, the new
approach is based on the variation of a trial hamiltonian
within a path integral framework and thus can also be
applied to relativistic and field theoretical problems.
We derive a diagrammatic scheme for the new approach and
show how a particular choice of the trial hamiltonian
corresponds exactly to the use of a Jastrow correlated
ansatz for the wave function in the Fermi Hypernetted Chain
approach. We show how our new approach can be used to
find upper bounds to ground state energies in systems
which the FHNC cannot handle, including those described
by an energy-dependent effective hamiltonian. We demonstrate
our approach by applying it to a quantum field theoretical
system of interacting pions and nucleons.
\end{abstract}
 
\pacs{21.60.-n;11.10.Ef;21.65.+f}


\narrowtext


\section{Introduction}

Variational methods provide extremely powerful tools for
analyzing complicated many-body
hamiltonians. Fermi Hypernetted Chain (FHNC) techniques, in
particular, have been successfully applied to many
different problems, ranging from nuclear matter to 
strongly interacting electrons to liquid helium, 
where strong correlations between particles dominate
the system and prohibit the usual perturbative
expansions \cite{FR74}-\cite{R82}. 
Correlated Basis Function (CBF) theory,
which employs FHNC and cluster summation methods
to compute perturbative
corrections to quantities determined variationally, has
also been successfully applied to a wide variety of
systems \cite{CMKKR79}-\cite{BFF91}. 
These methods, however, have up to now
been limited to hamiltonian systems for which a
nonrelativistic wave function can be defined. In the
present work we build upon a previous letter \cite{CF93},
developing a framework which goes beyond this limitation.
Our particular goal is to find a variational method
for studying nuclear systems which include dynamical
meson exchange. The methods we shall
develop, however, are much more general.

The basic idea of our approach is to adopt a 
Feynman path integral formulation in euclidean space 
so as to choose a trial hamiltonian which may
be determined variationally. This is less restrictive than the
usual FHNC approach, where one chooses a variational
wave function. The use of a variational principle in Feynman
path integrals is nothing new, as we describe in Sec. 2.
The novelty is connecting this principle to the extensive
existing FHNC calculations. This is a major goal of the present
work. We shall show that the FHNC diagrammatic expansion can
be recovered as a particular case of a more general
expansion based upon Feynman diagrams. Once this contact is
made, one may develop extentions of the highly successful
FHNC methods for summing up diagrams to be applied in the
more general path integral context, thus opening an
aspect of the path integral variational method
little explored up to now. Such techniques should have
a wide range of applications, including possible improvements
upon many existing FHNC calculations, as well as the analysis
of systems that the usual FHNC cannot handle, such as those
described by effective (time-dependent) hamiltonians.

The path integral formulation has another advantage:
it yields a perturbative expansion which is more compact, as
well as more general, than that of time-independent perturbation
theory. This can be a great technical advantage --- for
example, one Feynman diagram with $n$ interaction lines
corresponds in principle to $n!$ different Goldstone diagrams.
The path integral formulation thus provides a convenient setting
for the study of strongly interacting systems: the variational
principle can be used to find a trial hamiltonian which reproduces
as closely as possible the physics of the true hamiltonian, and
the corrections can be calculated within a natural perturbative
framework. Since it is usually a matter of some art to choose a
trial hamiltonian which successfully balances the competing
demands of accuracy and simplicity, a framework which allows a
ready interplay between the determination of a variational minimum
and perturbative corrections is clearly very useful.

The essential simplification which we shall adopt in choosing
a trial hamiltonian is motivated by the FHNC expansion. Whereas
a normal perturbative approach builds collective states from
independent particles, the FHNC instead treats interparticle correlations
as fundamental. Our idea is thus to choose a trial hamiltonian
$h_0$ containing a nontrivial two-body piece which may be
varied to reproduce as best as possible the interactions in the
true hamiltonian $h$, but we shall require the 
one-body part of $h_0$ to have a highly simplified form with
respect to that of $h$. In contrast to this, the usual perturbative
expansion takes $h_0$ to be just the one-body kinetic energy operator,
with no two-body term. Our choice, instead, gives the interactions
a more fundamental role; and the simplification of the one-body
part shall allow us, as in the FHNC expansion,
to sum entire classes of diagrams which give important contributions
in strongly correlated systems.

As a matter of fact, we shall choose a one-body piece which
depends on a single parameter $\alpha$, and we shall demonstrate
that the choice $\alpha=0$ (which corresponds to ``turning off''
the one-body piece completely)
exactly reproduces the FHNC expansion.
So one immediate result of our formulation is that we can
now calculate upper bounds on effective hamiltonians 
by using the full machinery of the usual FHNC approach.
The choice $\alpha=0$ is essentially a static approximation, and
to go beyond this approximation one must either take $\alpha>0$
or include higher order corrections in the perturbation $h-h_0$.
If we take $\alpha>0$, we must extend the FHNC techniques for
summing diagrams, because now the time integrations are no longer
trivial. While it is surely possible to use finite temperature
techniques to generalize the FHNC diagrammatic summation so
that $\alpha$ may be treated as a true variational parameter, in the
present work we shall be content with looking at the particular
case $\alpha\to\infty$. In this limit, the time integrations
simplify considerably, and we shall show a particular example of
how to sum up diagrams in analogy with the FHNC expansion.

For the particular case $\alpha=0$, the trial
hamiltonian $h_0$ is directly related
to a trial wave function of the CBF type, with the 
correlations in the CBF wave function corresponding
directly to the interactions in $h_0$.
Of course, as with the CBF correlations, there is a lot of freedom in
choosing the two-body part of $h_0$. Obviously, the closer one
can make this to that of $h$, the better one's variational estimate
of the energy will be.  The inclusion of state dependence (that is,
spin and isospin dependence) in $h_0$ clearly improves variational
estimates, although in the usual FHNC there are nevertheless
many difficulties in implementing this generalization. A popular
approximation is the so-called Single Operator
Chain approximation \cite{PW79}, which consists of summing
chains of single operator correlations without hyperconnecting
them. However, it seems necessary to go beyond the SOC \cite{WFF88}, 
and it is not yet well understood how to do this. A new 
approach using spin coherent-state wave functions currently
being developed \cite{FLSV96} is one possibilty.
While the present approach may very well provide an
alternative means of improvement
in this direction, either by allowing for a more complete summation
of state-dependent correlations through techniques developed for
nonzero $\alpha$ or by allowing for a more straightforward
calculation of state-dependent corrections through perturbation
theory, we shall limit ourselves in the following
to an $h_0$ which contains only spin-independent two-body interactions. 
This corresponds to using the
the Jastrow (state-independent) correlated trial wave function
ansatz. The study of how
best to include state-dependent correlations will thus be an
important immediate application of the approach we develop here.

The outline of the paper is as follows. We
describe the variational principle in the path integral
setting and discuss the choice of the trial hamiltonian in
Sections 2 and 3. Sections 4 through 6 are concerned with obtaining
the usual FHNC diagrammatic scheme as a particular limit of
a more general Feynman diagrammatic scheme. In the
process of demonstrating this, we show
that simplifications which allow the use of FHNC techniques for
summing diagrams do not occur in general. In Section 7, 
in a different limit from that which recovers
the FHNC diagrammatic scheme,  we
show how diagrams may be summed in
a different way, and hint at generalizations of the FHNC
techniques for this case. In Section 8 we demonstrate our
procedure by applying it to a nuclear system containing
dynamical pions. Finally, in Section 9 we discuss our
results and point toward directions for future study.

\section{The variational principle in terms of Feynman path integrals}

The original idea of formulating the variational principle in the path
integral context is due to Feynman himself \cite{F55,FH60}.  Let us
briefly summarize here the formalism.

As is well known, the ground state energy of a system can be expressed
in terms of the partition function according to 
\begin{equation}
E_0=-\lim_{\beta\to\infty}\frac{\log Z}{\beta}\;.  \label{mg1}
\end{equation} 
On the other hand, the partition function can be
expressed in a path integral form as 
\begin{equation} Z=\int
D[\psi^\dagger,\psi]e^{-\Ib
h[\psi^\dagger,\psi] d\tau},
\label{mg2} \end{equation} 
where we have assumed that the system can
be described in terms of a field $\psi$ (either fermionic or bosonic)
and its hermitian conjugate.  Here $ h[\psi^\dagger,\psi]$ is the
classical hamiltonian.  More precisely, when some ambiguities arise in
performing the L\'egendre transformation, we should start from the
lagrangian of the system $L[\psi^\dagger(t),\psi(t)]$ and make the
transformation to the euclidean world (imaginary time) $t\to it=\tau$.

Once a representation of $Z$ is provided, we can define an average
over the functional space where the Feynman-Ka\c c\ integral
(\ref{mg2}) is defined, namely, 
\begin{equation} 
\llan \; f\; \rran_{h_0}=\frac{1}{Z_0}\int
D[\psi^\dagger,\psi]e^{-\Ib
h_0[\psi^\dagger,\psi] d\tau}
f[\psi^\dagger,\psi]\; ,
\label{mg3}
\end{equation}
with
\begin{equation}
Z_0 = \int D[\psi^\dagger,\psi]
e^{-\Ib h_0[\psi^\dagger,\psi] d\tau }\;.  
\label{mg3.1} 
\end{equation} 
The weight function 
$e^{-\Ib h_0[\psi^\dagger,\psi]d\tau}$ must be
positive definite, which only implies that $h_0$
must be real.  Using this definition the partition function can be
rewritten in the form 
\begin{equation} 
Z=\int
D[\psi^\dagger,\psi]
e^{-\Ib h_0[\psi^\dagger,\psi] d\tau }
\llan e^{-\Ib [h-h_0]d\tau}\rran_{h_0}\;.  
\label{mg4}
\end{equation}
Now, it is a general property of a mean that 
\begin{equation}
\llan e^A\rran\geq e^{\llan A\rran}\;.  \label{mg5} 
\end{equation} 
Thus, applying this
last inequality to (\ref{mg4}) and inserting into (\ref{mg1}), we find
\begin{eqnarray}
\label{mg6} 
E_0 &\leq& \lim_{\beta\to \infty} \frac{1}{\beta}
\left[-\log \left(\int D[\psi^\dagger,\psi]e^{-\Ib
h_0 d\tau }\right)\right.\\
\nonumber && \qquad \left. + \llan\Ib h \, d\tau \rran_{h_0}
-\llan \int_{-\beta/2}^{\beta/2} h_0\, d\tau\rran_{h_0}\right]\;.  
\end{eqnarray}

The first term on the r.h.s. of
(\ref{mg6}) cancels with the third.  This outcome is
rather obvious if $h_0$ is time independent.  In this case the first
term is by definition the ground state energy of the hamiltonian
(being time independent and real, $h_0$ can in fact be regarded as a
hamiltonian) and the third term is the expectation value of the
same hamiltonian in the ground state, hence the cancellation.
However, since we are interested in extending the variational scheme
to effective hamiltonians and therefore want to maintain the
possibility of a time dependence in $h_0$, we need a more refined
proof, which we present in Appendix A. There we show that even
for time-dependent potentials, we end up with
\begin{equation} 
E_0\leq \lim_{\beta\to\infty}
\frac{1}{\beta} \llan \Ib  h(\tau)d\tau\rran_{h_0}\;.  \label{mg7}
\end{equation}

Reexamining the derivation of (\ref{mg7}), we see
that the functional integral describing $Z$ requires, in order to
ensure its existence, that $h$ must be real and bounded from below.  A
possible dependence upon the time in $h$ (or conversely, in
energy-momentum space, upon the energy) cannot be ruled out on the
ground of mathematical reasons.  We know that the hamiltonian is
energy-independent indeed, but it can be useful sometimes to derive an
effective hamiltonian, whose energy-dependence is the price we pay for
having eliminated some unwanted degrees of freedom.  Even in this case
the previous procedure applies safely with the only constraint that
the effective hamiltonian (obviously bounded from below because the
original one is) must be real in the euclidean world (that is, 
after the replacement $E \to iE$).

The same consideration applies to $h_0$ as well, which is nothing but
a weight function, even if it will be referred to in the following as
a ``trial hamiltonian":  again, the only condition $h_0$ must fulfill
for the variational principle (\ref{mg7}) to be mathematically
well-defined is that it must be real. Of course, for the 
$\beta\to\infty$ limit to be physically meaningful, we require that
$h_0$ must have an unique ground state. This is related to
the choice of boundary conditions in (\ref{mg2}), as we shall see in
Sec. 4. For the present we simply note that the inequality is valid
independent of the boundary conditions on the path integral, and indeed
also for finite $\beta$. Thus we have a variational principle
\begin{equation}
F_0\equiv -\log Z_0 + \llan \Ib [h-h_0]d\tau \rran_{h_0} \ge-\log Z
\end{equation}
for any $Z$ of the form
(\ref{mg2}), and the boundary conditions
and $\beta\to\infty$ limit must be chosen so as to make 
$F_0$ correspond to the quantity in which we are interested, namely,
the ground state energy.
 
Starting from (\ref{mg4}), we may also find perturbative
corrections in $h_I\equiv h - h_0$ to the variational
energy (\ref{mg7}) by simply expanding the
ratio
\begin{equation}
Z^\prime = \frac{Z}{Z_0} = \llan e^{-\Ib h_I d\tau}
\rran_{h_0}
\label{mg7.a}
\end{equation}
in Feynman diagrams. Because of the exponentiation property of
connected diagrams, we may write immediately
\begin{equation}
\label{7.b}
\ln Z^\prime = \llan e^{-\Ib h_I d\tau}
\rran_{h_0}^C = \sum_{n=1}^\infty \frac{(-1)^n}{n!}
\llan (\Ib h_I d\tau )^n \rran_{h_0}^C \; ,
\end{equation}
where $\langle...\rangle^C$ denotes connected diagrams.
The quantity
$W\equiv\ln Z=\ln Z_0+\ln Z^\prime$ 
is thus the sum of connected Feynman
diagrams containing no $h_I$ interaction lines, one
$h_I$ line, two $h_I$ lines, and so on. One
easily sees that to first order 
$-\lim\limits_{\beta\to\infty}W/\beta$ is the variational
energy (\ref{mg7}), and the higher orders bring in the
corrections. There will of course be a delicate interplay
between how far one wants to proceed in optimizing the trial
hamiltonian $h_0$ and how far one wants to be able to
go in calculating perturbative corrections, decisions
which will depend greatly upon the particular system
one is studying and upon the particular quantities
one is interested in calculating. We see here that the
Feynman path integral formulation provides a
convenient framework for arriving at such decisions.

\section{The trial hamiltonian}

Different choices of $h_0$ lead to different approximation
schemes.  For instance, we could choose $h_0$ in the form of a
single-particle kinetic energy
operator.  It is immediately seen that in such a case
the variational method coincides with the first order of the
usual perturbative expansion.  We could however assume $h_0$ in the
form of a single particle operator but retain a possible time
dependence.  This topic could give rise to interesting investigations,
because very much in the same line Feynman investigated the problem of
the polaron, obtaining numerical results far better than the ones
provided by the standard techniques (perturbation theory and canonical
transformations) \cite{F55}.

The problem can also be handled at a different level, by introducing
a more complicated trial hamiltonian having
not just a single particle but also a two
particle term.  The condition we require, of course, is that these
hamiltonians can be handled, technically, much more easily than the
full hamiltonian of the system.  Since a many body system is
intrinsically complicated by the presence of the two-body interaction,
no matter how simple this interaction is, the possibility we might
explore instead is to simplify as far as possible the single particle
term.
In this respect two options naturally arise, as we shall discuss in
the following.

Consider first a general lagrangian of a system in the real world,
namely 
\begin{eqnarray} &&L=i\int d^3x \dot\psi^\dagger(x)\psi(x)-\int
d^3x\, d^3x^\prime
\psi^\dagger(x)\epsilon(x,x^\prime)\psi(x^\prime)\nonumber\\
&&-\frac{1}{2}\int d^3x\, d^3x^\prime
\psi^\dagger(x)\psi^\dagger(x^\prime) U(|{\bf x}-{\bf
x}^\prime|)\psi(x^\prime)\psi(x) \label{mg7.1}, 
\end{eqnarray} 
where $\psi$ is taken from now on to represent a fermionic
field. Here we shall assume the system is translationally
invariant, so in 3-momentum space this becomes
\begin{equation} 
L=i\sum_{q}\dot a^\dagger_q
a_q-\sum_{q}\epsilon_qa^\dagger_q a_q-\tilde U \label{mg7.2}
\end{equation} 
where the $a$, $a^\dagger$ are the Fourier transforms
of the field operators, depending upon time and momentum, and with
$\tilde U$ we have denoted the Fourier transform of the potential part
of the lagrangian; $\epsilon_q$ is usually taken to be the 
single particle kinetic
energy.  Two obvious simplifications could be
to approximate this single particle 
energy either as zero or as a step function:
attractive below the Fermi sea and repulsive above.  These two cases
are intrinsically different, as we shall see in detail in the
following.  We can handle them together by setting 
\begin{equation}
\epsilon_q \longrightarrow \Delta
\theta(q-k_F)-\Delta \theta(k_F-q)\;;
\label{mg7.3} 
\end{equation} 
the former case is obtained from
(\ref{mg7.3}) by taking the limit $\Delta\to0$.

These two possible choices 
deserve further comment before proceeding.  The former has, as we shall
show, the great advantage of being able to reproduce
the FHNC diagrammatic scheme.  The latter instead
will provide the same diagrams although with different coefficients,
but nevertheless displays two advantages:  namely, it provides a
naturally well-defined ground state and furthermore the calculations,
when dynamical pions will be introduced, will turn out to be more
appropriate. This is because the latter case allows for a more
general choice of the two body term $\tilde U$.

For the sake of definiteness we write the lagrangian in the form
\begin{equation} 
L=\sum_{q}\dot a^\dagger_q
a_q-\Delta\sum_{q}\theta(q-k_F)a^\dagger_q a_q
+\Delta\sum_{q}\theta(k_F-q)a^\dagger_q a_q-\tilde U 
\label{mg7.6}
\end{equation} 
and its corresponding euclidean version
\begin{equation} 
L_E=L(t\to i\tau)\;, 
\label{mg7.7} 
\end{equation} 
and
then we identify $L_E$ with $h_0$.  We mention here that the
calculations, because of subtleties which we shall detail in
the next section,
are to be performed at finite $\beta$ and nonzero $\Delta$
and that only at the end of the calculation shall we decide in what
order we want to take the limits.

To summarize, the fundamental idea in the present scheme 
(and the same happens in the FHNC) 
is that the drastic assumption made on the single particle
part of $h_0$ greatly simplifies the diagrammatic expansion of
(\ref{mg7}) because all frequency integrations become trivial and the
four-dimensional problem reduces to a 3-dimensional one with constant
energy denominators.

Finally, we should mention that once we choose an $h_0$ that is
sufficiently complicated --- and the presence of a two-body term
ensures this --- then we are forced to make approximations in
evalutating the variational energy $E_0$. Of course, an approximation
to $E_0$ need not be an upper bound on the true ground state
energy; and thus we need to have a sufficiently robust
method for summing diagrams in an expansion in $h_0$ to allow
us to control the approximations made and to ensure that the
calculated energies converge systematically toward the true
variational energy $E_0$. The FHNC scheme has been successful in this
respect, at least in the case of spin-independent correlations,
and so we model our approach upon that scheme.

\section{Connection with the FHNC expansion}

The average value of a local, two-body
potential $V$ taken over a Jastrow correlated wave
function can be written as 
\begin{equation}
\langle \psi_J|V|\psi_J\rangle=
\langle \phi_0|e^{\U/2}Ve^{\U/2}|\phi_0\rangle
\label{mg8} 
\end{equation}
where $|\psi_J\rangle$ is the Jastrow-correlated approximation
to the ground state, whose wave function is 
\begin{equation}
\psi_J(r_1,\dots,r_N)=\hat{G}\phi_0(r_1,\dots,r_N)\;. 
\label{mg9}
\end{equation} 
Here $\hat G = e^{\U/2}$ is the Jastrow correlation operator
\begin{equation} 
\hat{G}=\prod_{ij}f(r_{ij})\;, 
\label{mg10}
\end{equation} 
$|\phi_0\rangle$ is the Free Fermi Gas ground state, and
$\phi_0(r_1,\dots,r_N)$ is the corresponding Slater determinant wave
function.

We wish to show that the same result comes out from the formula
\begin{eqnarray}  
\llan V\rran &=& \lim_{\beta\to\infty}
\label{mg11} 
\frac{1}{\beta}\int
D[\psi^\dagger,\psi]\\
&&\int d^4x d^4y
V(x-y)\psi^\dagger(x)\psi^\dagger(y)\psi(y)\psi(x)
e^{-\Ib d\tau \,L_E(\tau)} 
\nonumber 
\end{eqnarray}
where $ V(x-y)=V(|{\bf x}-{\bf y}|)\delta(x_0-y_0)$, provided a
suitable connection between $\cal U$ and $\tilde U$ is found.  At
first sight the connection is simple.  Since $V$ is 
time independent, we can replace (\ref{mg11}) with
\begin{eqnarray}  
\llan V\rran &=& \lim_{\beta\to\infty}
\label{mg11.1} 
\int
D[\psi^\dagger,\psi]
\int d^3x d^3y V(|{\bf x}-{\bf y}|)\\
&&\psi^\dagger({\bf x},0)\psi^\dagger({\bf y},0)
\psi({\bf y},0)\psi({\bf x},0)
e^{-\Ib d\tau \,L_E(\tau)},
\nonumber 
\end{eqnarray}
which can be rewritten, using the definition of
the path integral, as
\begin{equation}
\label{mg11.2}
\langle \phi_0 | {\cal T} \left( V(0) e^{-\Ib \, L_E d\tau}
\right)| \phi_0 \rangle
= \langle\phi_0| e^{-\int_0^{\beta/2}L_Ed\tau} V
e^{-\int_{-\beta/2}^0 L_E d\tau}|\phi_0\rangle
\end{equation}
and, since $U$ is also time independent, (\ref{mg11.2})
coincides with (\ref{mg8}) if we take
$L_E = -{\cal U}/\beta$, that is,
\begin{equation} 
U=-\frac{1}{\beta}\U\ \quad {\rm and}\quad \Delta = 0 \; .  
\label{mg13} 
\end{equation} 
Note that we have not invoked any commutation
properties between $U$ and $V$, and thus the expressions
(\ref{mg8}) and (\ref{mg11}) are also equivalent if
we replace $V$ by the kinetic energy $T$. These are
in fact particular instances of a more general
equivalence. For example, we could replace
$V$ in these equations by a nonlocal two-body
operator 
\begin{equation}
{\cal O} = \sum_{kpk^\prime p^\prime} O_{kpk^\prime p^\prime}
a^\dagger_k a^\dagger_p a_{p^\prime} a_{k^\prime},
\end{equation}
and the expressions would still be equivalent.
This implies, by means of functional derivation
with respect to $O$, that the two-body density
matrices derived from (\ref{mg8}) and (\ref{mg11})
also coincide.  It follows that the averages of
the kinetic energy from the Jastrow ansatz and
from the path integral approach also coincide.
We limit our discussion here to a local, two-body
operator for simplicity. 

In writing down (\ref{mg11.2}) we have
chosen a particular set of boundary conditions
for our path integral.
This shall be discussed presently, but it
is worthwhile here to first remind the reader of our
goal. The point is that (\ref{mg11}) is much more general
than (\ref{mg8}); in particular, there is no need to restrict the
potentials
$V$ or $U$ in (\ref{mg11}) to be independent of time, whereas this
is necessary if we use $V$ and $\U$ in (\ref{mg8})
(just as we call $h_0=L_E$ the trial
hamiltonian, we shall generally refer to the correlations $U$ and
${\cal U}$ as potentials). 
Thus our ultimate goal is
to develop a diagrammatic scheme for calculating (\ref{mg11}).
Here we are concerned with showing that in the particular case
when we choose a $V$ and a $U$ that can be handled by the
FHNC scheme arising from (\ref{mg8}), then for the choice
(\ref{mg13}) our new diagrammatic scheme will give exactly that
of the FHNC.

It is natural in this respect, starting from (\ref{mg11}), to
introduce a perturbation expansion in $U$.  First we define the
generating functional in the presence of external sources
\begin{equation} 
Z[\eta,\eta^\dagger]=\int
D[\psi^\dagger,\psi]e^{i \Ib 
\left[L(t)+\psi^\dagger(t)\eta(t)+\eta^\dagger(t)\psi(t) \right]dt}
\label{mg14} 
\end{equation} 
and its euclidean counterpart:  
\begin{equation}
Z_E[\eta,\eta^\dagger]=Z[\eta,\eta^\dagger]\Bigm|_{t\to i\tau},
\label{mg15} 
\end{equation} 
so that the average of the potential $V$ reads 
\begin{eqnarray} 
\lefteqn{\llan V\rran=} \label{mg16}\\ 
&&\int
d^4x\,d^4y V(x-y) \frac{\delta~~}{\delta \eta(x)}
\frac{\delta~~}{\delta \eta(y)}\frac{\delta~~}{\delta\eta^\dagger(y)}
\frac{\delta~~}{\delta\eta^\dagger(x)}
Z_E[\eta,\eta^\dagger]\Biggm|_{\eta,\eta^\dagger=0}\;.  
\nonumber
\end{eqnarray} 
Note that whether one uses a quantum-mechanical (real time)
or a statistical (imaginary time)
approach is irrelevant in determining a static quantity
(a real number) like the average of $V$.  Thus we can use
indifferently the generating functional or the partition function in
the following, at least as long as we are interested in static (ground
state) properties of the system.  The study of linear responses should
instead be dealt with only in the quantum-mechanical frame.

The perturbation expansion is now carried out on the generating
functional by means of the relation 
\begin{equation}
Z[\eta,\eta^\dagger]=e^{-i\int
d^4x\,d^4y\,U(x-y)\frac{\delta~~}{\delta \eta(x)}
\frac{\delta~~}{\delta \eta(y)}\frac{\delta~~}{\delta\eta^\dagger(y)}
\frac{\delta~~}{\delta\eta^\dagger(x)}} Z_0[\eta,\eta^\dagger]
\label{mg17} 
\end{equation} 
where $U(x-y)=U(|{\bf x}-{\bf y}|)\delta(x_0-y_0)$, 
the ``unperturbed'' lagrangian is 
$L_0=L+\tilde U$, with $L$ given in (\ref{mg7.6}), and 
\begin{equation}
\label{mg18}
Z_0[\eta,\eta^\dagger]=
\int D[\psi^\dagger,\psi]e^{i\Ib 
\left[L_0(t)+\psi^\dagger(t)\eta(t)+
\eta^\dagger(t)\psi(t) \right]dt}\; .
\end{equation}

The reason we have repeated here these straightforward manipulations
is that they point out two subtleties of the procedure.

The first subtlety is that the boundary conditions are exploited in
explicitly evaluating the path integral in (\ref{mg18}),
that is, in defining the inverse of the lagrangian $L_0$.
This free propagator in momentum space reads 
\begin{equation}
\G{\omega_n}{q}=\frac{\theta(q-k_F)}{\omega_n-\Delta+i\eta}+
\frac{\theta(k_F-q)}{\omega_n +\Delta -i\eta} 
\label{mg19}
\end{equation} 
with discrete frequencies because the time integration
is limited:  
\begin{equation} 
\omega_n=\frac{n\pi}{\beta}\;.
\label{mg20} 
\end{equation} 
Here we wish to choose antiperiodic boundary conditions in
order that the path integral is equivalent to a trace
over fermionic states, and so $n$ is restricted 
to be an odd integer.

The second subtlety, strictly related to the first,
concerns the degeneracy of the unperturbed ground state:  as
long as $\Delta$ is kept finite an unique ground state exists 
and coincides
with the Fermi sphere, and the definition (\ref{mg19}) makes sense both
for finite $\beta$ and in the limit $\beta\to\infty$.  Then a
perturbation theory can be performed by means of eq.  (\ref{mg17}).
If we put instead $\Delta=0$ from the very beginning, any independent
particle state is degenerate with respect to $L_0$ and fixing the
ground state becomes arbitrary.  Thus, taking the limit
$\beta\to\infty$ for $\Delta\not=0$ picks out the ground state when
the boundary conditions on the path integral are chosen 
as above, that is, so that
(\ref{mg18}) gives a trace over states.
For $\Delta=0$, however, we must fix the
ground state ``by hand'' --- that is, as is detailed in
Appendix B, we must choose a more 
involved set of boundary conditions which explicitly give
the states $\phi_0$ at $t=-\beta/2$ and $\phi^\dagger_0$ at $t=\beta/2$
in (\ref{mg17}).  Moreover, with $\Delta=0$
the momentum space propagator (\ref{mg19}) is then singular in the
$\beta\to\infty$ limit.

To demonstrate that we are truly evaluating the average of $V$
as given in (\ref{mg8}) we need the expression of the propagator in
configuration space, namely:
\begin{eqnarray}
\G{x}{y}&=& 
\langle \phi_0|{\cal T}\left(\psi(x)\psi^\dagger(y)\right)|\phi_0\rangle
\label{mg21}\\
&=&e^{-i\Delta|x_0-y_0|} 
\intt{k}e^{i {\bf k} \cdot ({\bf x} - {\bf y})}\nonumber \\
&&\qquad\left[
\theta(x_0-y_0)\theta(k-k_F)-\theta(y_0-x_0)\theta(k_F-k)\right]\;.
\nonumber 
\end{eqnarray}
This corresponds to the second choice of boundary
conditions discussed above, so that we can properly take the $\Delta\to 0$
limit. Thus the Fourier transform of (\ref{mg21}) is equal to
(\ref{mg19}) only in the $\beta\to\infty$ limit (which must
be taken in the end, in any case).
 
With some elementary manipulations and
defining the matrix element of the free density matrix in
configuration space as 
\begin{equation} 
\rho(k_F|{\bf x}-{\bf y}|)=
\langle{\bf x}|\rho|{\bf y}\rangle
=\intt{k}e^{i {\bf k}\cdot ({\bf x}-{\bf y})} \theta(k_F-|{\bf k}|)\;, 
\label{mg22}
\end{equation} 
we can rewrite (\ref{mg21}) in the form
\begin{equation}
\G{x}{y}=e^{-i\Delta|x_0-y_0|}\left\{\theta(x_0-y_0)\delta({\bf
x}-{\bf y})-\rho(k_F|{\bf x}-{\bf y}|)\right\}\;.  
\label{mg22.1}
\end{equation} 
In the euclidean world, one has  
\begin{equation}
\GE{x}{y}=e^{-\Delta|x_0-y_0|}\left\{\theta(x_0-y_0)\delta({\bf
x}-{\bf y})-\rho(k_F|{\bf x}-{\bf y}|)\right\}\;.  
\label{mg23}
\end{equation} 
To complete the proof that this perturbative expansion
leads to the FHNC diagrams it is now sufficient to follow the
derivation of the FHNC diagrams given in \cite{BR89}:  there it
is shown that the FHNC diagrams are obtained from a Feynman diagram
expansion followed by time integration (we shall show this in
some detail in Sec. 6).  The identification (\ref{mg13}) provides
the link between the potentials, and the free Green's function is
nothing but $\GE{x}{y}\bigm|_{\Delta=0}$.

Thus we can conclude that the FHNC 
diagrammatic scheme exactly follows from
the perturbation expansion (\ref{mg17}) in the limit
$\Delta\to0$ followed by the limit $\beta\to\infty$.  
We stress that it need
not be equivalent --- and this is indeed the case --- if one
exchanges the order of these limits.  We can infer the reason for
this trouble.  It is apparent in fact from (\ref{mg23}) that two
cutoffs are present, the first being $\beta$ and the second $\Delta$,
and dropping the first or the second need not necessarily give the
same results.  It will be proved in the next section that the two
expansions obtained by taking the limits in different orders are in
fact not equivalent.

\section{The order of the limits $\beta\to\infty$ and $\Delta\to0$}

The fundamental task in constructing the FHNC diagrams is to perform,
in some way, the time integrations in the Feynman diagrams, in order
to get a static or 3-dimensional theory.  As a first step we prove, as
a byproduct, that the two limits we discussed before are indeed not
equivalent.  We consider for simplicity in this section only vacuum to
vacuum diagrams, but of course the procedure can be easily generalized
to any kind of Green's function.

To start with, let us consider a generic Feynman diagram of order $n$
(that is, a diagram that contains $n$ potential lines).  It will
contain, of course, $2n$ fermionic lines ${\cal G}^0_\Delta$, which we
write in the form (\ref{mg21}), in which the different
time orderings are explicit.  We shall work
in configuration space, and we are now interested in carrying out the
time integrations.  The general form of the diagram will be 
\begin{equation} 
\frac{1}{\beta}\int
d^3x_1\dots d^3x_n\int\limits_{-\frac{\beta}{2}}^{\frac{\beta}{2}}
d\tau_1\dots \int\limits_{-\frac{\beta}{2}}^{\frac{\beta}{2}}d\tau_n
\prod \G{x_i}{x_j} \prod U(|{\bf x}_k-{\bf x}_l|) ,
\label{mg24}
\end{equation} 
where we have divided by $\beta$ in order to take the
$\beta\to\infty$ limit.

As a first step we can break the time integration region in $n!$ parts,
each one characterized by a definite ordering of the times.  For a
given ordering in eq.  (\ref{mg24}), then, only one of the two terms
composing the Green's function (\ref{mg21}) can contribute.  So each
line is characterized by its propagation forward or backward in time,
and the $n!$ terms so obtained are, in general, different ---
although, in practice, many are equivalent.  We will refer to them 
--- improperly --- as Goldstone diagrams.  We observe that all 
Goldstone diagrams derived from the same Feynman diagram have the 
same spatial part, and it is only the time integrations 
which are, in principle, different.

Having chosen a Feynman diagram and having expanded it in Goldstone
diagrams, let us further select one of these diagrams and
perform its time integrals.  The diagram is characterized by $n$ time
variables, in a given time ordering.  Let them be
$\tau_1<\tau_2<\dots<\tau_n$.  The $\theta$-functions will then be
absorbed into the integration limits.  The time attenuation factors
$e^{-\Delta|\tau_i-\tau_j|}$ remain, but now all absolute values can
be removed.  It can be easily seen that in any diagram, for each
potential line taken at a time $\tau$, each propagator connected to
the potential line from below (no matter if incoming or outgoing)
carries a factor $e^{-\Delta\tau}$ and each propagator connected from
above carries a factor $e^{\Delta\tau}$.  Now let us make a trick.
For each fermion line crossing but not stopping at a given time
$\tau_i$ let us insert the factors $e^{\Delta\tau_i}
e^{-\Delta\tau_i}$.  Then let us collect together all the factors
$e^{-\Delta(\tau_{i+1}-\tau_i)}$.  Their number is just the number of
fermion lines flowing between $\tau_i$ and $\tau_{i+1}$ (including,
then, also those which do not end at $\tau_i$ or $\tau_{i+1}$).  Thus
let us call $n_1$ the number of fermion lines flowing between $\tau_1$
and $\tau_2$ and, more generally, $n_i$ those between $\tau_i$ and
$\tau_{i+1}$.  The time part of the diagram will then be
\begin{equation}
I(\beta,\Delta) = 
\frac{1}{\beta}\int\limits_{-\frac{\beta}{2}}^{\frac{\beta}{2}}d\tau_n
\int\limits_{-\frac{\beta}{2}}^{\tau_n}d\tau_{n-1}\dots\int\limits_{
-\frac{\beta}{2}}^{\tau_3}d\tau_2\int\limits_{-\frac{\beta}{2}}^{\tau_
2} d\tau_1 \prod_{i=1}^{n-1} e^{-n_i\Delta(\tau_{i+1}-\tau_i)}\; .
\label{mg25} 
\end{equation} 
Now consider the two different limits:
\begin{enumerate} 
\item $\Delta\to0$

In this case the integral is trivial:  we get 
\begin{equation}
I(\beta,\Delta) = \frac{\beta^{n-1}}{n!}\;.
\label{mg26} 
\end{equation} 
\item $\beta\to\infty$

Changing variables to $\lambda_i=\tau_i-\tau_{i+1}$ and
$\tau_n\to\beta\tau_n$, we find
\begin{equation}
\lim_{\beta\to\infty} I(\beta,\Delta) =
\int\limits_{-1/2}^{1/2}d\tau_n\int\limits_{-\infty}^{0}
d\lambda_{n-1}\dots\int\limits_{-\infty}^{0}d\lambda_1
\prod_{i=1}^{n-1}e^{n_i\Delta\lambda_i}=
\prod_{i=1}^{n-1}\frac{1}{n_i\Delta}\; . 
\label{mg28} 
\end{equation}  
\end{enumerate}

We are now ready to prove the inequivalence of the two limits.

Consider first the diagram of fig.  \ref{fig1}:  it generates 6
Goldstone diagrams, all equal and all having the same graphical
representation as in fig.  \ref{fig1}.  The time integrations are
immediate according to our preceding rules and, accounting for the
multiplicity of the diagram, they provide a factor $\beta^2$ for the
limit $\Delta\to0$ and $\frac{6}{16\Delta^2}$ for the limit
$\beta\to\infty$.

Next consider the diagram of fig.  \ref{fig2}.  It generates 24
Goldstone diagrams:  this time, however, not all are equivalent.  As a
matter of fact, three kinds of diagrams arise, each with multiplicity
8, as shown in fig.  \ref{fig3}.  Their evaluation is now simple
according to our rules.  For the sum, we find $\beta^3$ in the
limit $\Delta\to0$, of course, and 
$$8\frac{1}{(4\Delta)^3}+8\frac{1}{(4\Delta)^3}+
8\frac{1}{(4\Delta)^2(8\Delta)}=\frac{5}{16\Delta^3}$$
for the limit $\beta\to\infty$.  It is thus clear that no linear
relationship between $\Delta$ and $\beta$ can be found which is able
to provide the same result for the two graphs.  This counterexample
proves the inequivalence between the different orderings of the
two limits.

In fact, it is clear from (\ref{mg24}) that by re-scaling
$U\to U/\beta$, the time integrations give a factor
${\tilde I}(\alpha)\equiv I(\beta,\Delta)/\beta^{n-1}$, that is,
a function of just the product $\alpha\equiv\beta\Delta$.
Thus we could send $\beta\to\infty$ and $\Delta\to 0$ in
such a way that $\alpha$ is finite, and then treat $\alpha$
as a new variational parameter. One can see from (\ref{mg25})
that the time integrations are then in general nontrivial.
Of course, the results (\ref{mg26}) and (\ref{mg28}) correspond
to $\alpha\to 0$ and $\alpha\to\infty$, respectively; and, because
the time integrations simplify greatly in these limits,
these are the only cases we shall consider in the following.

\section{Derivation of the FHNC diagrams}

To carry out the time integrations and to derive the FHNC diagrams we
start from the separation (\ref{mg22.1}).  It is convenient to
introduce new symbols in the diagrams.  Thus let us set
\begin{equation} 
\K(x;\beta)=e^{-i\Delta|x_0|}\theta(x_0)\delta(x)\;,
\label{mg29} 
\end{equation} 
graphically denoted from now on with a
double solid line, and 
\begin{equation}
\g(x;\beta)=-e^{-i\Delta|x_0|}\rho(k_F |{\bf x}|)\;, 
\label{mg30}
\end{equation} 
denoted by a single line, so that ${\cal
G}^0_\Delta={\cal K}^0_\Delta+g^0_\Delta$.  The procedure we shall
follow is now, from a given Feynman diagram, to expand it into
products of $\K$ and $\g$, thus generating a certain number of new
diagrams --- which we shall call ``intermediate diagrams'' ---
with single and double lines, as exemplified in fig.  \ref{fig4}.
Then we shall re-sum classes of these new diagrams in order to
simplify the subsequent time integrations.  In fact, for the case
$\Delta=0$, we can sum together the intermediate diagrams in such a
way as to eliminate all time dependence in the fermion lines, so that
the time integrations are trivial and the resulting diagrams are
equivalent to those of the FHNC expansion. 

In order to reproduce the FHNC, of course, we consider potentials
$U$ (corresponding to Jastrow correlations) which are independent of
time.  Thus the only time dependence in the diagrams we are
considering comes from the fermion propagators.  (For the case
$\Delta$=0, in particular, we see from (\ref{mg29}) and
(\ref{mg30}) that the only time dependence is in
the double line propagators $\K$.)  Thus, if we can demonstrate that 
we can sum up intermediate diagrams in such a way that --- without ever
performing any integrations --- time dependence can be eliminated in
an arbitrary line of fermion propagators, then we will have succeeded
in demonstrating that {\it all} the intermediate diagrams can be summed to
form a new set of diagrams which are entirely time
independent.  Again, for the case $\Delta=0$,
it is in fact sufficient to consider a fermion line
consisting of an arbitrary number of double-line propagators preceded
and followed by a single line propagator.  We shall begin with the
simplest case.

To get rid of the time dependence of the intermediate diagrams, we
want to cancel the $\theta$-functions carried by the double lines.
To this purpose consider first a double line in a generic diagram
preceded and followed by a single line, as in fig.  \ref{fig5}a.  Then
another diagram differing from the first only in
the subdiagram of fig.  \ref{fig5}b surely also exists, independent of
what is contained in the rest of the diagram.
The subdiagrams plotted in fig.  \ref{fig5} may be summed
immediately to provide (in Euclidean space, and with the obvious
notational shortcuts $\rho_{ij}=\rho(k_F|{\bf x}_i-{\bf x}_j |) =
\rho_{ji}$ and $\delta_{ij}=\delta(|x_i-x_j|)$) 
\begin{eqnarray}
\nonumber && \delta_{12}\rho_{13}\rho_{14}
e^{-\Delta|t_1-t_2|}\Big[e^{-\Delta|t_1-t_3|}
e^{-\Delta|t_2-t_4|}\theta(t_2-t_1)\\ 
&& \qquad\qquad\qquad
+e^{-\Delta|t_2-t_3|} e^{-\Delta|t_1-t_4|}\theta(t_1-t_2)\Big]\;.
\label{mg30.1} 
\end{eqnarray} 
Two points in this equation deserve
comment:  first, the spatial part of the diagram is factorized, owing
to the spatial $\delta$-function contained in the double line
propagator and, second, the time structure is complicated and depends
explicitly upon the time ordering, as one can easily verify.  It
clearly simplifies drastically in the limit $\Delta\to 0$, however,
when it becomes just $\delta_{12}\rho_{13}\rho_{14}$.  
In this limit, then, the sum of the subdiagrams in fig. 5
is a new, time-independent subdiagram whose spatial 
part is identical to that of these two subdiagrams.
 
The next step is to
consider the subdiagrams of fig.~\ref{fig6}.  
It is easily realized here
that, always in the limit $\Delta\to0$, again all the spatial
parts factor out and the time part becomes a sum of
$\theta$-functions, just the sum of all possible time orderings, and
it gives exactly 1.

The procedure can be iterated in this way for any
number of subsequent double lines.

Other particular cases are straightforward.  
For instance, if the two external
lines join at the same point (i.e., $t_3=t_4$) clearly nothing
changes, and the proof follows the same path if the external lines are
replaced by a single external line closing the loop (it is sufficient
to multiply the diagram by $\left[g^0_\Delta\right]^{-1}(x_4-x_3)$ and
to integrate over $t_3$ and $t_4$).  Since any closed loop must
contain at least one single line (because only the single line pieces
carry an advanced part) we have proved that, collecting a suitable
class of Feynman diagrams together, in any closed loop the
$\theta$-functions of ${\cal K}^0_\Delta$ cancel out.
Since now any
closed diagram contains closed loops only, our proof holds true for
all closed diagrams.  In this case all time dependence disappears, and
the time integrations are trivial, giving a factor $\beta^n$.  Note
that this proof does not depend upon the particular form of the
dashed (potential) lines.  They represent, in general, the
two-body part of the trial hamiltonian $h_0$, namely $U=-{\cal
U}/\beta$ (the factors $1\over\beta$ cancel
the $\beta$'s coming from the time integrations), but any one line
in a given diagram should
be replaced by $V$, namely, the true potential of the
system.  In this way the closed diagrams just represent the average
value of $V$ in terms of time-independent diagrams.

Before concluding the discussion of the diagrammatic scheme in the limit
$\Delta\to0,$ we want to stress a last detail.  It is not always
straightforward to determine which diagrams must be collected together
in order that the $\theta$-functions sum to 1.  The needed diagrams
surely exist, {\em but they could also coincide}.  That is, a given
symmetry could mean that $n$!  of the necessary diagrams might not
correspond to distinct Feynman diagrams.  In this case we simply get a
factor $1/n!$ in front of the diagram.  This is in particular the case
for the ladder diagrams of fig.  \ref{fig7}.  
There, clearly,
accounting for the factor $1/n!$ and for the locality of the
propagators ${\cal K}^0_\Delta$, the sum reads 
\begin{equation} 
{\cal U}(|{\bf x}-{\bf y}|)+\frac{1}{2}{\cal U}^2(|{\bf x}-{\bf y}
|)+\frac{1}{3!}  {\cal U}^3({|\bf x}-{\bf y}|)+\dots=e^{\cal U}-1\;.
\label{mg31} 
\end{equation} 
This provides the still unspecified link
between $U = -{1\over\beta} {\cal U}$ and the variational parameter
$f(r_{ij})$ used in the FHNC.  There, in fact, no ladder diagrams
are present, but those of fig.  7 can be summed up to give the result
(\ref{mg31}), so that the desired link is 
\begin{equation} 
f^2 -1 = e^{\cal U}-1\;.  
\label{mg32} 
\end{equation}

Having thus summed together the various classes of intermediate
diagrams to obtain a new set of time-independent diagrams, we may
shrink the double-line propagators --- which now just represent spatial
$\delta$-functions --- to points.  We may furthermore replace each
potential line with the ladder sum $e^{\cal U}-1$ above.  Although
more than one potential line can meet at a single point, we must now
specify that any two given points can be connected by at most
only one potential line so as to
avoid overcounting.  There is one spatial propagator $\rho_{ij}$
entering and one leaving each point (including the possibility 
$\rho_{ii}$), 
and all diagrams contain only
closed fermion loops.  This completes the derivation of the FHNC
diagrammatic scheme.

Let us now make the following observations. First, it is clear
that for $\Delta\neq 0$ our re-summing of intermediate diagrams does
not eliminate all time dependence, and we will need to look for a
modification of FHNC techniques in order to handle the nontrivial
time integrations in this case. Second, we should
like to stress once again that our derivation of the FHNC diagrams in
the case $\Delta=0$ is entirely independent of the form of the
potentials. In particular, the unlabelled ends of the potential
lines in figs. 5 and 6 could be fixed at different times without
disrupting the conclusions that the fermion lines in the re-summed
intermediate diagrams are time independent. Thus our approach allows
for the possibility that the true potential $V$ and/or the correlations
$U$ can be time dependent. Because the fermion lines are nevertheless
time independent in this latter case, the time integrations of
an $n$th order diagram become simply independent averages of the $n$
potentials over time. Consequently, for the case $\Delta=0$,
there will be no gain in allowing the correlations to depend on
time; nevertheless, our formalism now allows us to find variational
upper bounds on the ground state energies of
time-dependent effective hamiltonians, and
no changes in the FHNC machinery
are needed to implement this extension. We shall see a specific
example of this latter case in sec. 8.

\section{The diagrams for finite $\Delta$}

Consider now the case of the limit $\beta\to\infty$ followed
by $\Delta\to0$.
This time let us make the rescaling $U\to \Delta U$ and consider any
diagram of order $U^{N-1}$ in the expansion of the average of $V$.
After the frequency integrations in momentum space, it is easy to see
that there will be $N-1$ energy denominators which are just multiples
of $\Delta$ (this can also be seen from our configuration space
analysis in sec.  4).  Since the diagram also carries $N-1$ $U$
lines, after the rescaling all the $\Delta$'s cancel so that the
result is independent of $\Delta$.  Thus the limit $\Delta\to0$ is
unnecessary. This is, in fact, just a re-statement of our
conclusion from sec. 5: namely, that it is only the product
$\alpha=\Delta\beta$ which is important here. The FHNC diagrammology
is reproduced by taking the $\alpha\to 0$ limit, and the present
case corresponds to the limit $\alpha\to\infty$.

Apart from the different rescaling, the derivation of the diagrams
follows precisely the same steps as in the previous section but,
having separated the $\K$ and the $\g$ lines, the frequency
integration provides, as seen before, different coefficients.  Once
these have been found by explicit integration, then all $\K$ lines
disappear and, up to the different coefficients, we are left with
exactly the same diagrams as before.  This time, however, the
re-summation of the class of diagrams of fig.~\ref{fig7} follows a
different path.  Let us work in momentum space and let us define
$\tilde U(q)$ as the Fourier transformed potential and
$\tilde\Gamma(p,k;q)$ as the sum of the ladder diagrams, $p$ and $k$
being the momenta entering the diagrams.  Clearly
$\tilde\Gamma(p,k;q)$ fulfills the Bethe-Salpeter equation
\begin{eqnarray} 
\tilde\Gamma(p,k;q)&=&\tilde
U(q)+i\intq{t}{\tilde\Gamma(p,k;t)\tilde U(q-t)
\over\left(p_0+t_0-\Delta+i\eta\right)\left(k_0-t_0-\Delta+i\eta\right
)} \nonumber\\ &=&\tilde
U(q)+\frac{1}{k_0+p_0-2\Delta+i\eta}\intt{t}\tilde\Gamma(p,k;t)\tilde
U(q-t)\;.  
\label{mg33} 
\end{eqnarray} 
Going to configuration space,
this equation is immediately solved.  Calling $x_1$, $x_2$ and
$y_1-y_2$ the transforms of $p$, $k$ and $q$, we get
\begin{eqnarray}
\label{mg34} 
\Gamma(x_1^0,x_2^0,y_1-y_2)&=&
\int\limits_{-\infty}^{\infty}\frac{dp_0}{2\pi}
\int\limits_{-\infty}^{\infty}\frac{dk_0}{2\pi}
e^{ip_0(x_1^0-y_1^0)+ik_0(x_2^0-y_2^0)}\\ 
&&\qquad
\frac{U({\bf y_1-y_2})}
{ 1-\frac{\strut\displaystyle U({\bf
y_1-y_2})}{\strut\displaystyle
k_0+p_0-2\Delta+i\eta}}\delta(y_1^0-y_2^0) 
\nonumber
\end{eqnarray}
which still depends upon the initial times.

However, we recall that {\em no conditions have been imposed upon the
possible time (or energy) dependence of $U$}.  In particular, we can
rewrite the trial hamiltonian (now, strictly, no longer a hamiltonian)
as 
\begin{equation} 
h_0=\sum_{k>k_F}\Delta a_k^\dagger a_k -
\sum_{k<k_F}\Delta a_k^\dagger a_k
+\Delta\sum_{kpq}\tilde\lambda(k,p;{\bf q})a_p^\dagger a_k^\dagger
a_{k-q} a_{p+q} 
\label{mg35} 
\end{equation} 
where we explicitly allow
an energy dependence in the ``effective potential'' $\tilde\lambda$
upon the initial energies $k_0$ and $p_0$ but not upon the transferred
energy $q_0$.  We can now write down the Bethe-Salpeter 
equation exactly as
before, getting a similar result.  Going again to configuration space
and calling $\lambda$ the Fourier transform of $\tilde\lambda$ with
respect to the spatial coordinates only, we get 
\begin{eqnarray}
\label{mg35.1}
\Gamma(x_1,x_2,y_1,y_2)&=&
\int\limits_{-\infty}^{\infty}\frac{dp_0}{2\pi}
\int\limits_{-\infty}^{\infty}\frac{dk_0}{2\pi}
e^{ip_0(x_1^0-y_1^0)+ik_0(x_2^0-y_2^0)}\\
&& \qquad
\frac{\lambda(p_0,k_0,{\bf y_1-y_2})}{ 1-\frac{\strut\displaystyle
\lambda(p_0,k_0,{\bf y_1-y_2})}{\strut\displaystyle
k_0+p_0-2\Delta+i\eta}}\delta(y_1^0-y_2^0)\; .  
\nonumber
\end{eqnarray}
Now we can use the extra energy dependence of $\lambda$
to remove that of $\Gamma$.  We can define, in fact, 
\begin{equation}
\lambda(p_0,k_0,{\bf y})=\frac{\left(p_0+k_0-2\Delta\right) 
\gamma({\bf y})} {p_0+k_0-2\Delta+\gamma({\bf y})} \label{mg36}
\end{equation} 
so that inserting (\ref{mg36}) into (\ref{mg35}) we get
the straightforward result 
\begin{equation}
\Gamma(x_1,x_2,y_1,y_2)=\gamma({\bf
y_1-y_2})\delta(x_1^0-y_1^0)\delta(x_2^0-y_2^0) \delta(y_1^0-y_2^0)\;.
\label{mg37} 
\end{equation} 
Thus, having chosen a
particular frequency dependence of the effective potential $\tilde
\lambda$, we can substitute everywhere the potential with the sum of
the ladder series $\Gamma$, which is now fully frequency independent.
Of course at this stage the only relevant point is the existence of
the starting quantity $\tilde \lambda$, not its cumbersome form.
Naturally, the contact with the FHNC diagrammatic scheme is now provided by
$f^2-1=\gamma$.  

However, we must remind the reader that in the
$\beta\to\infty$ limit in which we are working now, the coefficients
of our diagrams resulting from frequency (or time) integration are not
those of the FHNC because the fermion lines are not time
independent.  We still have to make the connection with FHNC
techniques for summing diagrams in this case, for the different
coefficients prevent us from summing diagrams through the
usual FHNC integral equations. While we shall not pursue 
this topic further in the present work, we will mention that 
this $\Delta
\neq 0$ case might be handled by extending the FHNC equations
to include time integrals, thereby generating the different
coefficients as one builds chains and hyperchains. 
It is convenient to work
in configuration space, but to transform from time to frequency
variables. Then one arrives at a diagrammatic scheme which is
identical to that of the FHNC except that exchange and dynamical
correlation lines carry frequency variables, and the points --- while
still representing spatial $\delta$-functions --- now have
a nontrivial frequency dependence. Because of the
simple form of the single particle term in $L_0$, however,
one can still
classify diagrammatic types in a way similar to that
done in the FHNC, forming
chains by performing convolution integrals over frequency as
well as spatial variables, and forming hyperchains by summing
over ladder diagrams as demonstrated here for the simplest type
of two-point correlation.
An important extention
here would be to allow for correlations ${\tilde\lambda}$ in
(\ref{mg35}) which also depend on the frequency $q_0$.

\section{The pion exchange potential}

Finally we come to a quantum field theoretical model in order to see
how it can be embedded in the previous formalism.  Consider for the
sake of simplicity a system of non-relativistic nucleons interacting
with a pionic field through the typical interaction lagrangian
\begin{equation} 
L_I=-i\frac{f_{\pi NN}}{m_\pi}\int d^3x
\psi^\dagger(x)\left(\bsig\cdot \bfm\nabla \vec\tau \cdot
\vec\phi(x)\right)\psi(x)\;.  
\label{mg38} 
\end{equation} 
If $L_N$ and
$L_\pi$ are the free lagrangians for nucleons and pions, respectively,
the partition function of the system reads 
\begin{equation} 
Z=\int
D[\psi^\dagger\psi\phi]e^{-\Ib d\tau
\left\{L_N+L_\pi +L_I\right\}_{t\to i\tau}} 
\label{mg39}
\end{equation} 
where it must be stressed --- and this is the main
drawback of the present model --- that the term $\lambda\phi^4$ in the
lagrangian has been neglected, whereas we know that it is necessarily
present in a renormalizable theory.  This simplification ensures,
however, that the remaining lagrangian is quadratic in the pion
field, which can consequently be integrated out.  Calculations are
straightforward \cite{ACMS87} and provide (apart from an
inessential constant factor) 
\begin{eqnarray} 
\lefteqn{Z=\int
D[\psi^\dagger\psi]} \label{mg40}\\ 
&&e^{-\Ib dx^0
L_N -\Ib dx^0
\Ib dy^0
\int d^3x d^3y
\bfm{\scriptstyle\nabla}_x \cdot {\rm\bf j}_F^a(x)
D_0(x-y) \bfm{\scriptstyle\nabla}_y \cdot {\rm\bf j}_F^a(y)
\bigm|_{t \to i\tau}} 
\nonumber 
\end{eqnarray} 
where the fermion current is
\begin{equation}
{\rm\bf j}_F^a(x) = -i
\psi^\dagger(x)(\bsig \tau^a )\psi(x)
\end{equation}
and
$D_0(x-y)$ denotes the free pion propagator, which in momentum space
and in the euclidean world reads 
\begin{equation} 
\tilde
D_0(q,\omega_n)=\frac{1}{\omega^2_n+q^2+m_\pi^2} 
\label{mg41}
\end{equation} 
with $\omega_n$ given of course by (\ref{mg20}), this time
with $n$ even integer only.  We can then follow all our machinery to
derive the average value of the interaction term, which looks exactly
as before if we disregard its frequency (or time) dependence.

Now from (\ref{mg16}) we observe that the diagrams we need to 
evaluate contain one pion
propagator (hence a line with the extrema put at different times)
closed by a two-particle Green's function whose time evolution is
governed by the trial hamiltonian.  Let us now take the limit
$\Delta\to0$, $\beta\to \infty$ ($\alpha\to 0$)
as in sec.~6.  We recall
that the rules for obtaining the FHNC diagrams derived there are not
influenced by a possible time dependence in the interaction lines.
More explicitly, it is irrelevant if the potential lines in figs.
\ref{fig5} and \ref{fig6} are instantaneous or not.  In both cases,
for a given Feynman diagram, others exist (including, perhaps,
equivalent diagrams) such that when these are added together the
$\theta$-functions sum up to 1, and all time dependence drops out.
Consequently the two-particle Green's function evaluated in the frame
of FHNC turns out to be time-independent.  Let this be, say,
$G_{II}({\bf x_1}, {\bf x_2},{\bf x_3},{\bf x_4})$.
The integral we are interested in is then 
\begin{eqnarray} 
\lefteqn{\frac{1}{\beta}\Tr\int d^3x\,
d^3y \Ib dx_0\Ib dy_0}
\label{mg42}\\ 
&&D_E(x-y)\left(-i\bsig\cdot
\bfm\nabla_x\vec\tau\right) \left(-i\bsig\cdot
\bfm\nabla_y\vec\tau\right) G_{II}({\bf x},{\bf y},{\bf x},{\bf y})\;.
\nonumber 
\end{eqnarray} 
Now we carry out the integrations over the
imaginary times.  Using eq.~(\ref{mg41}), it is a simple matter to
show that 
\begin{eqnarray} 
\lefteqn{\frac{1}{\beta}
\Ib dx_0\Ib dy_0 D_E(x-y) }
\label{mg43}\\ 
&&=\frac{1}{\beta}
\Ib dx_0\Ib dy_0
\int\frac{d^3q}{(2\pi)^3} e^{-i{\bf q\cdot(x-y)}} \sum_{n\; even}
\frac{e^{i\omega_n(x^0-y^0)}}{\omega_n^2 + {\bf q}^2 + m_\pi^2}
\nonumber\\ 
&& = \int\frac{d^3q}{(2\pi)^3} e^{-i{\bf
q\cdot(x-y)}}\frac{1}{{\bf q}^2 + m_\pi^2} = 
\frac {e^{-m_\pi |{\bf x}-{\bf y}|}}{4\pi |{\bf x}-{\bf y}|}\;.  
\nonumber 
\end{eqnarray} 
The conclusion is that what we need for
the calculation is to take the effective potential at zero energy and
then to perform the standard FHNC calculations.  In the present model, the
effective potential is just the OPEP, as one might have expected.

To go beyond this variational approximation, now, we have several 
possibilities. We can try to improve the variational result by
allowing the trial hamiltonian to have a spin-dependent interaction
and/or a time dependence by taking $\Delta\ne0$. Otherwise, we can
keep the simpler trial hamiltonian used here, which allows the use
of the full FHNC machinery, and we can calculate corrections to
the variational result through perturbation thery as described
in sec. 2.  Implementation of these improvements is currently
underway.

\section{Conclusions and Outlook}

Let us now briefly summarize the results obtained in this
paper and discuss the possible developments they naturally
suggest.

The central idea in this paper is that of formulating 
a variational approach for dealing with strongly interacting
systems which goes beyond the limitations of the Fermi
Hypernetted Chain and Correlated Basis Function methods. 
This led us to use a path integral approach, where  the variational
method consists in choosing a trial hamiltonian instead of choosing
an ansatz for a nonrelativistic wave function.
We have obtained several important results. First, one can
define the path integral and choose the trial hamiltonian
in such a way that for particular systems for which 
a nonrelativistic wave function makes sense, our approach
produces exactly the same results as the
successful FHNC formulation. Second,
in this path integral formalism one is
not linked to a true hamiltonian, but an
effective hamiltonian can be used as well. Thus ultimately we can generalize
the FHNC variational techniques for use in studying
the enormous variety of strongly correlated systems
for which the usual FHNC cannot be directly applied
because a formulation in terms of nonrelativistic wave functions
breaks down. Thirdly, we have shown how corrections to a static
variational approximation can be added by choosing a more
general trial hamiltonian so to improve the variational energy
and by including perturbative corrections in the difference
between the true and trial hamiltonians within a compact
path integral formulation.

Having first briefly described the variational principle in the
Feynman path integral formulation, we proceeded to define
our approach by using the FHNC and CBF approaches as motivation.
Since these  latter approaches treat correlations between
particles as more important than independent particle  kinetic
energies, in our approach the fundamental simplification 
with respect to the true hamiltonian $h$ we chose for our trial
hamiltonian $h_0$ was to modify its one-body piece.
For much of the paper we limited ourselves to the case of a time-independent
$h_0$, and throughout we only considered spin-independent interactions
in $h_0$. We made a specific simple choice of the one-body term
in $h_0$ which depends on a parameter $\alpha = \beta \Delta$, 
and we have shown that in the particular case
$\alpha \to 0$ our expansion scheme in terms of Feynman diagrams
exactly reproduces the FHNC results for a variational wave function
of the Jastrow type. We also showed that in the limit $\alpha \to
\infty$, the time integrals are also simple, if not as trivial as in
the case $\alpha=0$. We showed that this other limit
corresponds to a new approximation scheme which has the same 
diagrammatic structure as the FHNC except with
different coefficients in front of each
diagram.

Finally, to show how our method works in practice, we have shown that the
variational scheme can be applied to a quantum-field-theoretical case,
a system of interacting nucleons and pions. We have proved a minimal
but highly nontrivial result --- namely, that in a system
containing dynamical pions a variational procedure built up upon the static
one pion exchange potential still provides an upper bound to the energy.

The present paper was mostly devoted to proving the correctness of the method
and its equivalence, in a very particular case, with the available FHNC
calculations. However, it opens several interesting perspectives which we want to
discuss now.
First, we limited ourselves to trial hamiltonians which correspond to
the Jastrow ansatz for the wave function in the usual FHNC approach. 
State-dependent interactions in $h_0$ can be handled on the same footing
 as those described here.
Very likely, as some preliminary results seem to indicate, the method 
presented here could provide a very useful tool
to overcome the limitations of the Single Operator Chain approximation.

Next, we have to exploit the new possibilities offered by the present 
approach. 
First, energy-dependent correlations can be accounted for, thus improving
upon the usual calculations carried out with phenomenological nuclear
potentials. 
Second, an extension to relativistic systems is now allowed, potentially
cumbersome but in principle feasible. In fact, 
again we are faced, in a relativistic
system, with a case where a trial wave function looses its sense but
our (eventually time-dependent) trial hamiltonian does not.
Third, in connection with the previous point, the possibility of carrying out
a variational calculation with the Bonn potential \cite{MHE87}
is now offered. Note
that using energy-dependent interactions in the trial hamiltonian 
(energy-dependent correlations in the FHNC language)
allows for an improvement upon the results
of sec. 8, going beyond the static limit of the potential.

Finally, another promising application of our approach is to quantum
field theory, as indeed we briefly demonstrated in sec. 8. 
Variational methods
have been little used in field theory up to now (see, for example,
\cite{VQFT}-\cite{H91}), and to our knowledge nothing of
the sort introduced here
to deal with strongly correlated systems has been explored in the
field theoretical problems. 
Our approach thus opens up new possibilities in this
direction. The obvious, if ambitious, application
would be to study low energy bound states of quantum chromodynamics,
where the usual perturbative methods are hopeless. There are two
paths one can follow in this respect. One is to study euclidean
path integrals as in sec. 8, obtaining an effective lagrangian 
through functional integration. A second possibility, which
might overcome some of the problems mentioned by Feynman \cite{F88},
 is first to
develop an effective hamiltonian using Wilsonian renormalization
group methods --- a promising approach in this direction 
is proposed and detailed in \cite{WW94} --- and then to
analyze this effective, frequency-dependent hamiltonian with our
path integral variational methods. Indeed, it was the availability
of powerful hamiltonian methods such as the variational principle 
which helped lead Wilson to develop his renormalization group
approach \cite{W65}. In either of the approaches to field theoretical
systems mentioned here, of course,
one has difficult renormalization problems to overcome.

It is our hope, then, that the variational approach described
here will eventually provide an important tool for studying
many different types of strongly correlated systems for which
the usual perturbative techniques are insufficient, and that
only a small number of the potential applications of our
approach have been mentioned here.
\vskip1cm

\centerline{\bf Acknowledgements}

We acknowledge helpful discussions with F. Conte,
P. Saracco and E. Migli.

\appendix

\section{Derivation of eq. (8)}

We show here that the first and third terms on the r.h.s. of
eq. (\ref{mg6}) cancel even for a time-dependent potential
$h(\tau)$.

Let us define a suitable ``partition function'' 
\begin{equation}
Z_0^\lambda=\int D[\psi^\dagger\psi]e^{-\lambda\Ib
h_0(\tau)d\tau}\;,  \label{mgA1} 
\end{equation} 
so that
\begin{equation} 
-\frac{\partial~}{\partial\lambda}\log Z_0^\lambda=
\frac{\int D[\psi^\dagger\psi]
\left( \Ib h_0(\tau)d\tau \right)
e^{-\lambda\Ib h_0(\tau)d\tau}}{\int
D[\psi^\dagger\psi]e^{-\lambda\Ib
h_0(\tau)d\tau}}\;.  \label{mgA2} 
\end{equation} 
By comparison with (\ref{mg3})
\begin{equation} 
\llan \Ib h_0(\tau)d\tau)\rran_{h_0}= -
\frac{\partial~}{\partial\lambda}\log Z_0^\lambda\Bigm|_{\lambda=1}\;.
\label{mgA3} 
\end{equation} 
Now we represent $h_0(\tau)$ in
(\ref{mgA2}) as the Fourier series $\sum_{\omega_n} e^{i\omega_n
\tau} {\tilde h_0}(\omega_n)$ with $\omega_n = \pi n/\beta$, 
so that as $\beta\to\infty$
\begin{equation}
\Ib h_0(\tau) d\tau \longrightarrow \beta {\tilde h_0}(0).
\end{equation}
In this limit, then, the path integral is equivalent to a
trace over the eigenstates of ${\tilde h_0}(0)$,
whose eigenvalues we write as ${\tilde\epsilon}_n$. Thus
\begin{eqnarray}
-\frac{1}{\beta}\frac{\partial~}{\partial\lambda} \log
Z_0^\lambda \label{mgA4.1} &\longrightarrow& 
\frac{\int D[\psi^\dagger\psi]{\tilde h_0}(0) 
e^{-\lambda\beta {\tilde h_0}(0)}}{
\int D[\psi^\dagger\psi]e^{-\lambda\beta {\tilde h_0}(0)}}\\
&\longrightarrow& \frac{\sum_n
{\tilde \epsilon_n} e^{-\lambda\beta {\tilde \epsilon_n}}}{
\sum_n e^{-\lambda\beta {\tilde \epsilon_n}}}\;.
\nonumber 
\end{eqnarray}
As $\beta\to\infty$, of course, the lowest energy state
dominates this average, giving the result 
$$-\lim_{\beta\to\infty}\frac{1}{\beta}
\frac{\partial~}{\partial\lambda}\log Z_0^\lambda
= {\tilde\epsilon_0},$$ which does not depend
upon $\lambda$.  This implies 
\begin{equation}
\lim_{\beta\to\infty}\frac{1}{\beta}
\frac{\partial~}{\partial\lambda}\log Z_0^\lambda =
\int\limits_{0}^{1}d\lambda\lim_{\beta\to\infty}\frac{1}{\beta}
\frac{\partial~}{\partial\lambda}\log
Z_0^\lambda=\lim_{\beta\to\infty} \frac{1}{\beta} \log
Z_0^\lambda\Bigm|_{\lambda=1} \label{mgA4} 
\end{equation} 
since $Z$ is normalized such that
$\log Z_0^\lambda\Bigm|_{\lambda=0}=0$. Comparing (\ref{mg3}),
(\ref{mgA3}) and (\ref{mgA4}), we end up with 
\begin{equation} 
E_0\leq \lim_{\beta\to\infty}
\frac{1}{\beta} \llan \Ib  h\,d\tau\rran_{h_0}\;.  \label{mgA5}
\end{equation}

\section{Boundary conditions on the path integral}

Here we show explicitly the two choices of boundary conditions
on the ferm-ionic euclidean path integral
discussed in Sec. 4. Further
details may be found, for example, in Ref. \cite{BR89}.

The standard notation
\begin{equation}
Z = \int D[\psi^\dagger,\psi] e^{-\Ib h_0[\psi^\dagger,\psi]d\tau}
\label{mgB.0}
\end{equation}
used in the text is a misleading shorthand for
\begin{equation}
Z(z_f^*, z_i) = 
\int\limits^{z^*(\beta/2)=z_f^*}_{z(-\beta/2)=z_i}
\lim_{N\to\infty}\prod_{n=-N}^{N}
 d\mu[z({\beta n/2N})] e^{-\Ib h_0(z,z^*)d\tau},
\label{mgB.1}
\end{equation}
where the integration measure is
\begin{equation}
d\mu[z] = \prod_q dz^*_q dz_q e^{-z^*_q z_q}.
\end{equation}
Here $h_0$ is assumed time-independent for simplicity, and
the classical Grassman variables $z$ and $z^*$ correspond
to the nonorthogonal coherent states 
\begin{equation}
| z \rangle = e^{\sum_q z_q a_q^\dagger}|0\rangle \quad
{\rm and}\quad \langle z | = \langle 0| e^{\sum_q 
z_q^* a_q},
\end{equation}
with $|0\rangle$ the Fock vacuum. These coherent
states are nonorthogonal superpositions of Fock states with differing
numbers of particles, and their overlap is
\begin{equation}
\langle z | z^\prime \rangle = e^{\sum_q z_q^* z_q^\prime}
\equiv e^{z^* z^\prime}.
\end{equation}
The variables $z_i$ and $z_f^*$ are not integrated over
in (\ref{mgB.1}),  which is equivalent to the matrix element
\begin{equation}
\langle z_f | e^{-\beta h_0} | z_i \rangle.
\end{equation}

The usual choice of boundary conditions
is to set $z_f^* = -z_i^*$ and to integrate over $z_i$
\begin{equation}
Z = \int d\mu(z_i) Z(-z_i^*,z_i),
\end{equation} 
so that the generating
function $Z$ becomes a trace over all Fock states --- that
is, the partition function. Then the
$\beta\to\infty$ limit will pick out the ground state, provided
$h_0$ has a unique ground state. If this is not the case, as
for the FHNC choice $\Delta=0$, we must work a little harder
to ensure that (\ref{mgB.1}) is equivalent to the desired
matrix element. We can use the closure relation
\begin{equation}
\int d\mu[z] |z\rangle \langle z| = 1
\end{equation}
to represent any state $|\phi\rangle$ as
\begin{equation}
|\phi\rangle = \int d\mu[z] |z \rangle \langle z|\phi\rangle\;.
\end{equation}
Thus in the $\Delta=0$ case we must take our shorthand expression
(\ref{mgB.0}) to represent
\begin{equation}
Z = \int d\mu[z_f]d\mu[z_i] \langle \phi_0 | z_f\rangle
Z(z_f^*, z_i) \langle z_i | \phi_0 \rangle
\end{equation}
in order to ensure that we are evaluating the matrix
element $\langle \phi_0 | e^{-\beta h_0} | \phi_0 \rangle$.
Here $|\phi_0\rangle$ is the free Fermi gas ground state as
in Sec. 4, so that
\begin{equation}
\langle \phi_0|z\rangle = \prod_{q\le k_F} z_q.
\end{equation}

\newpage
 
\begin{figure}
\caption{Third order Feynman diagram} 
\label{fig1}
\end{figure}

\begin{figure}
\caption{Fourth order Feynman diagram} 
\label{fig2}
\end{figure} 

\begin{figure}
\caption{Time-ordered ``Goldstone'' 
diagrams corresponding to fig. 2} 
\label{fig3}
\end{figure}

\begin{figure}
\caption{Example of separation 
of propagators into two terms} 
\label{fig4}
\end{figure}
 
\begin{figure}
\caption{Two subdiagrams 
which sum to give a time-independent
diagrammatic structure}
\label{fig5}
\end{figure}
 
\begin{figure}
\caption{ A further example of summation to form 
time-independent diagrams}
\label{fig6}
\end{figure}

\begin{figure}
\caption{ Summation of ladder diagrams}
\label{fig7}
\end{figure}

\end{document}